\documentclass[aip,php,twocolumn,reprint,showpacs,superscriptaddress,groupedaddress,showkeys,amssymb,floatfix]{revtex4-1}

\usepackage{bm} 
\usepackage{verbatim}    
\usepackage{mathptmx}    
\usepackage[colorlinks=true,allcolors=blue]{hyperref} 
\usepackage{graphicx}    
\usepackage{subfigure}   
\usepackage{lipsum}
\usepackage{float}
\usepackage[none]{hyphenat}
\usepackage[T1]{fontenc}
\usepackage{color}
\usepackage{soul}

\begin{document}
\title{Sustained neutron production from a sheared-flow stabilized $Z$-pinch}

\author{Y. Zhang}
\email[]{yzhang16@uw.edu}
\affiliation{Aerospace \& Energetics Research Program, University of Washington, Seattle, WA 98195 USA}

\author{U. Shumlak}
\affiliation{Aerospace \& Energetics Research Program, University of Washington, Seattle, WA 98195 USA}

\author{B. A. Nelson}
\affiliation{Aerospace \& Energetics Research Program, University of Washington, Seattle, WA 98195 USA}

\author{R. P. Golingo}
\affiliation{Aerospace \& Energetics Research Program, University of Washington, Seattle, WA 98195 USA}

\author{T. R. Weber}
\affiliation{Aerospace \& Energetics Research Program, University of Washington, Seattle, WA 98195 USA}

\author{A. D. Stepanov}
\affiliation{Aerospace \& Energetics Research Program, University of Washington, Seattle, WA 98195 USA}

\author{E. L. Claveau}
\affiliation{Aerospace \& Energetics Research Program, University of Washington, Seattle, WA 98195 USA}

\author{E. G. Forbes}
\affiliation{Aerospace \& Energetics Research Program, University of Washington, Seattle, WA 98195 USA}

\author{Z. T. Draper}
\affiliation{Aerospace \& Energetics Research Program, University of Washington, Seattle, WA 98195 USA}

\author{J. M. Mitrani}
\affiliation{Lawrence Livermore National Laboratory, Livermore, CA 94550 USA}

\author{H. S. McLean}
\affiliation{Lawrence Livermore National Laboratory, Livermore, CA 94550 USA}

\author{K. K. Tummel}
\affiliation{Lawrence Livermore National Laboratory, Livermore, CA 94550 USA}

\author{D. P. Higginson}
\affiliation{Lawrence Livermore National Laboratory, Livermore, CA 94550 USA}

\author{C. M. Cooper}
\affiliation{Lawrence Livermore National Laboratory, Livermore, CA 94550 USA}

\date{\today}



\begin{abstract}
The sheared-flow stabilized (SFS) $Z$-pinch has demonstrated long-lived plasmas with fusion-relevant parameters. This Letter presents the first experimental results demonstrating sustained, quasi-steady-state neutron production from the Fusion $Z$-pinch Experiment (FuZE), operated with a mixture of 20\% deuterium/80\% hydrogen by volume. Neutron emissions lasting approximately $5~\mu$s are reproducibly observed with pinch currents of approximately $200$~kA during an approximately $16~\mu$s period of plasma quiescence. The average neutron yield is estimated to be $\left ( 1.25\pm 0.45 \right )\times 10^{5}$ neutrons/pulse and scales with the square of the deuterium concentration. Coincident with the neutron signal, plasma temperatures of $1-2$~keV, and densities of approximately $10^{17}$~cm$^{-3}$ with $0.3$~cm pinch radii are measured with fully-integrated diagnostics.
\end{abstract}
\maketitle

Over the years, fusion neutron production has been investigated on multiple experiments reporting promising results: solid-liner compression of field-reversed configuration (FRC) experiments,\cite{taccetti2003frx,intrator2004high} laser-driven magnetic-flux compression (LDFC) experiments at the OMEGA laser facility,\cite{gotchev2008magneto,gotchev2009seeding} magnetized liner inertial fusion (MagLIF),\cite{slutz2010pulsed,slutz2012high} and plasma liner driven magneto-inertial fusion (MIF).\cite{hsu2012spherically,hsu2012experimental} The $Z$-pinch is another well-known concept capable of generating fusion neutrons.\cite{post1956controlled, haines1981dense} Although the first observations of deuterium-deuterium fusion neutrons from $Z$-pinches were reported in the 1950s,\cite{kurchatov1957possibility,berglund1957fusion,anderson1958neutron} classical magnetohydrodynamic (MHD) instabilities severely limited the realization of high performance fusion plasmas.\cite{haines2011review,velikovich2007z} However, theoretical and computational investigations illustrate that sufficient radial-shear of an axial flow can stabilize $Z$-pinches against $m$ = 0 and $m$
= 1 MHD modes.\cite{shu1995sheared,terry2000suppression,paraschiv2010linear} The role of axial flow shear has been examined experimentally in the ZaP and ZaP-HD (high density) experiments with reported pinch currents of approximately 50 kA.\cite{shumlak2001evidence,shumlak2003sheared,shumlak2012sheared,shumlak2017increasing} The fundamental result of these studies is that the $Z$-pinch can be stabilized by applying a sheared axial flow, $dV_z / dr$, above the criterion $0.1kV_{A}$ where $k$ is the axial wave number and $V_{A}$ is the Alfv\'{e}n velocity.\cite{shu1995sheared,shumlak2001evidence}

Recently the scaling of the SFS $Z$-pinch concept towards fusion conditions is investigated at the FuZE facility. This Letter presents the demonstration of the SFS $Z$-pinch at approximately four times higher pinch current with fusion-relevant plasma parameters and the first evidence of fusion neutron generation from an SFS $Z$-pinch. Stabilized $Z$-pinch plasmas with densities of approximately $10^{17}$~cm$^{-3}$, ion temperatures of approximately $1-2$~keV, and pinch radii of $0.3$~cm are achieved. Sustained, quasi-steady-state neutron production is observed for approximately $5~\mu$s during the $16~\mu$s quiescent period, coincident with pinch currents of approximately $200$~kA. Yields of approximately $10^5$ neutrons/pulse are detected.

\begin{figure*}[htb]
\includegraphics[width=\textwidth]{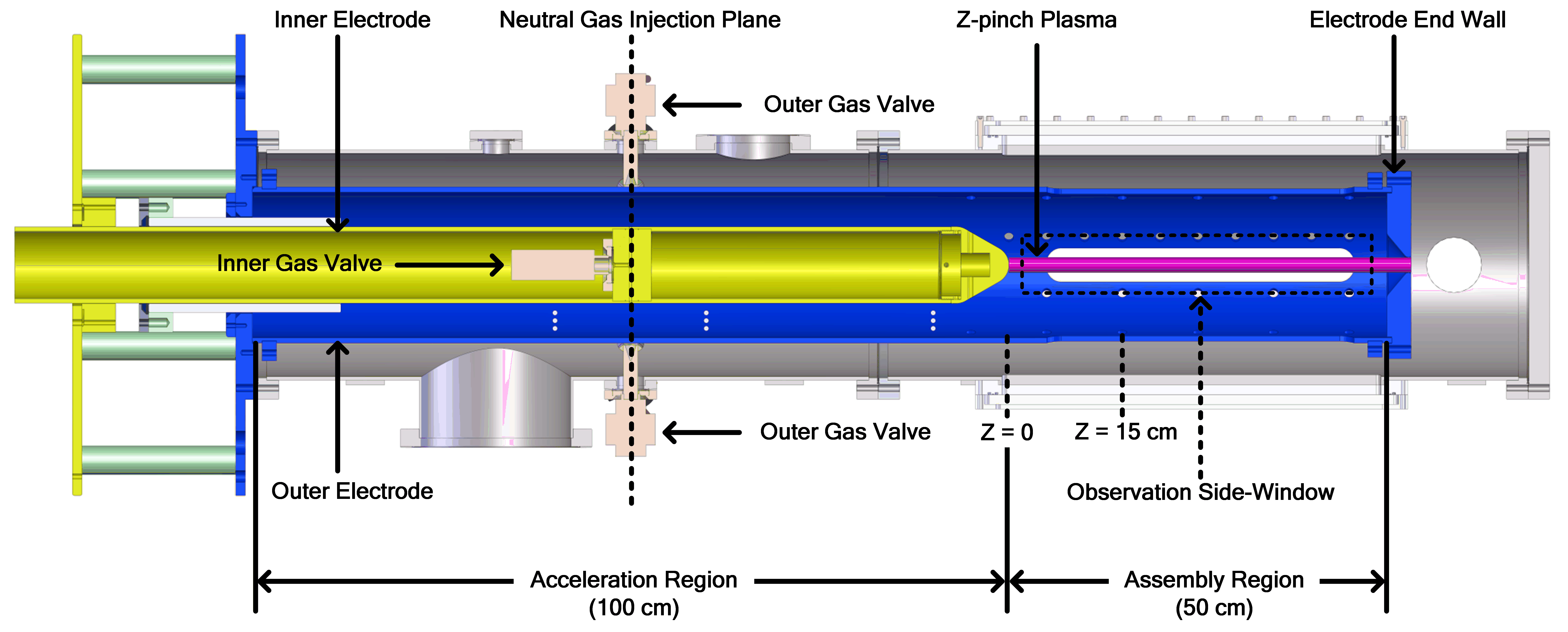}
\caption{Side-view of FuZE. The diameters of the coaxial inner and outer electrodes are $10$~cm and $20$~cm, respectively. Multiple axial arrays of surface-mounted magnetic probes measure the azimuthal magnetic field and mode amplitude. A digital holographic interferometer measures plasma density from a side view of the plasma. The plasma temperature is measured by Doppler broadening via impurity ion emission spectroscopy from a top view of the plasma. Neutron yields are measured by a cylindrical plastic scintillator (Eljen, EJ-204), coupled to a photomultiplier tube (PMT). The experimental data presented here are obtained at the $z = 15$~cm plane.}
\label{fig:overview}
\end{figure*}

Figure~\ref{fig:overview} shows a schematic depiction of the FuZE device. A $100$~cm coaxial acceleration region is coupled to a $50$~cm pinch assembly region. A shaped copper nose cone is attached at the exit of the acceleration region. The assembly region is formed by extending the outer electrode (blue in Fig.~\ref{fig:overview}) $50$~cm beyond the end of the inner electrode (yellow in Fig.~\ref{fig:overview}). The working gas is puffed into the acceleration region at $z = -50$ cm through one gas puff valve injecting radially through the inner electrode and four gas puff valves injecting radially through the outer electrode, as labeled in Fig.~\ref{fig:overview}. Attached to the end of the outer electrode is a copper electrode end wall. The outer electrode contains four slots for optical diagnostic access, as shown in Fig.~\ref{fig:overview}.

\begin{figure}[htb]
\includegraphics[width=0.98\columnwidth{}]{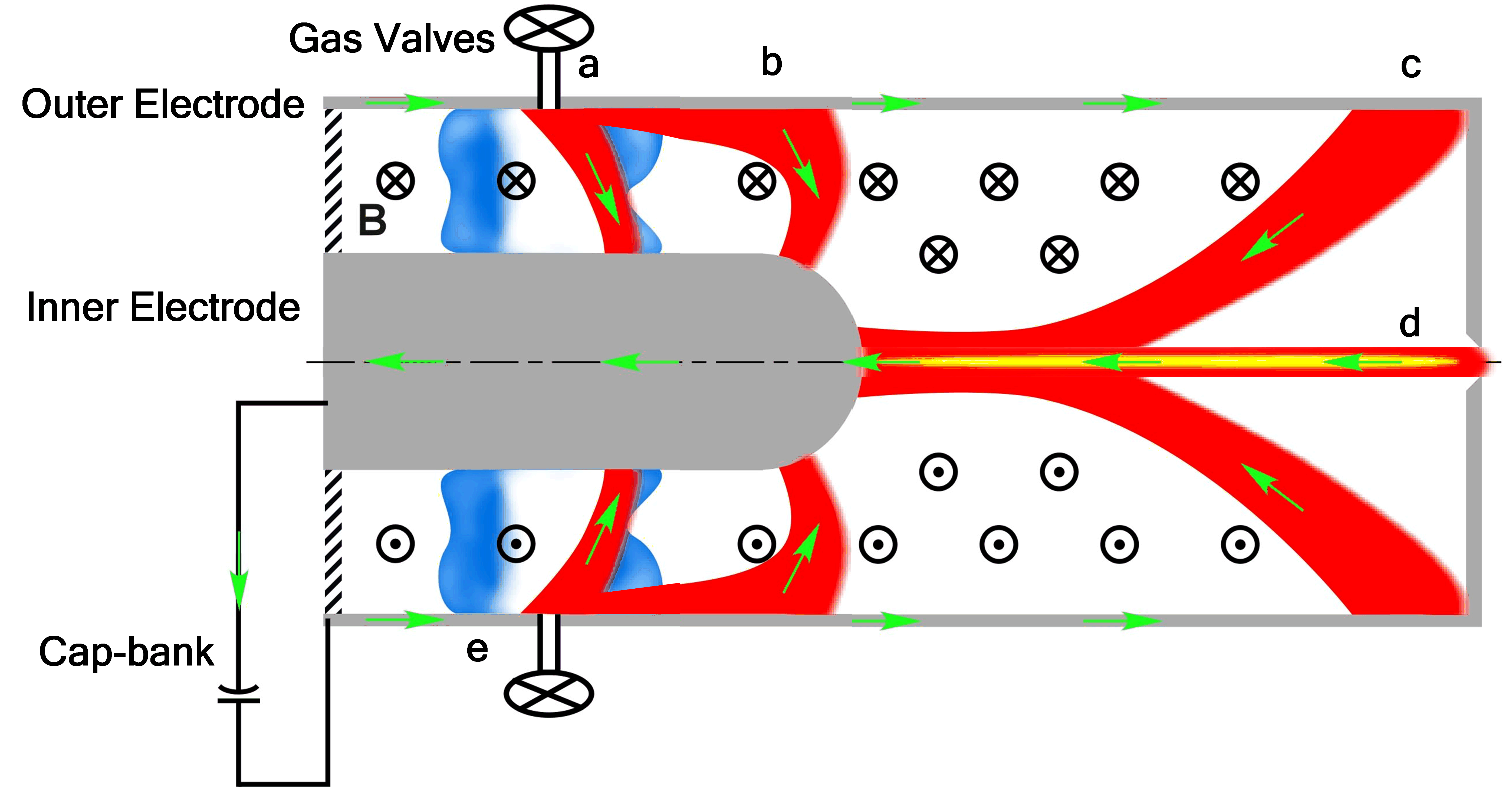}
\caption{A simplified schematic of an SFS Z-pinch plasma formation showing an overlay of five times (a--e). (a) Neutral gas (blue) is injected into the annular acceleration region and then ionized. (b) The {\bf J}$\times${\bf B} force (driven current and resulting magnetic field) accelerates the plasma (red) axially along the coaxial accelerator. (c) At the end of the accelerator, the plasma transitions from the inner electrode to the axis. (d) The $Z$-pinch plasma forms in the assembly region. (e) A deflagration process supplies continuous plasma flow into the assembly region. The current is indicated with green arrows.}
\label{fig:principle}
\end{figure}

A schematic illustrating how the FuZE device creates an SFS $Z$-pinch plasma is shown in Fig.~\ref{fig:principle}. During this dynamic process, an inherently-generated axial-flow-shear stabilizes the $Z$-pinch plasma and initiates the quiescent period. A deflagration process in the acceleration region maintains the sheared axial flow and the resulting quiescent plasma equilibrium.\cite{loebner2016radial} Eventually the depletion of plasma supplied from the acceleration region leads to the decrease of the assembly region plasma density, the increase of Alfv\'{e}n velocity, and the end of quiescent period. Detailed discussion on SFS $Z$-pinch plasma quiescence has been reported elsewhere.\cite{shumlak2009equilibrium}

A suite of diagnostics, including holography, spectroscopy, magnetic probes, and plastic scintillator detectors, is used for characterizing plasma properties and for addressing the physics of neutron production. The experimental measurements presented in this Letter are taken from over 150 pulses with the same experimental conditions. Pulses consistently show similar pinch current, quiescent period of plasma, and neutron emission pulse behavior.

\begin{figure}[htb]
\centering
\subfigure[] {\includegraphics[width=0.505\columnwidth{}]{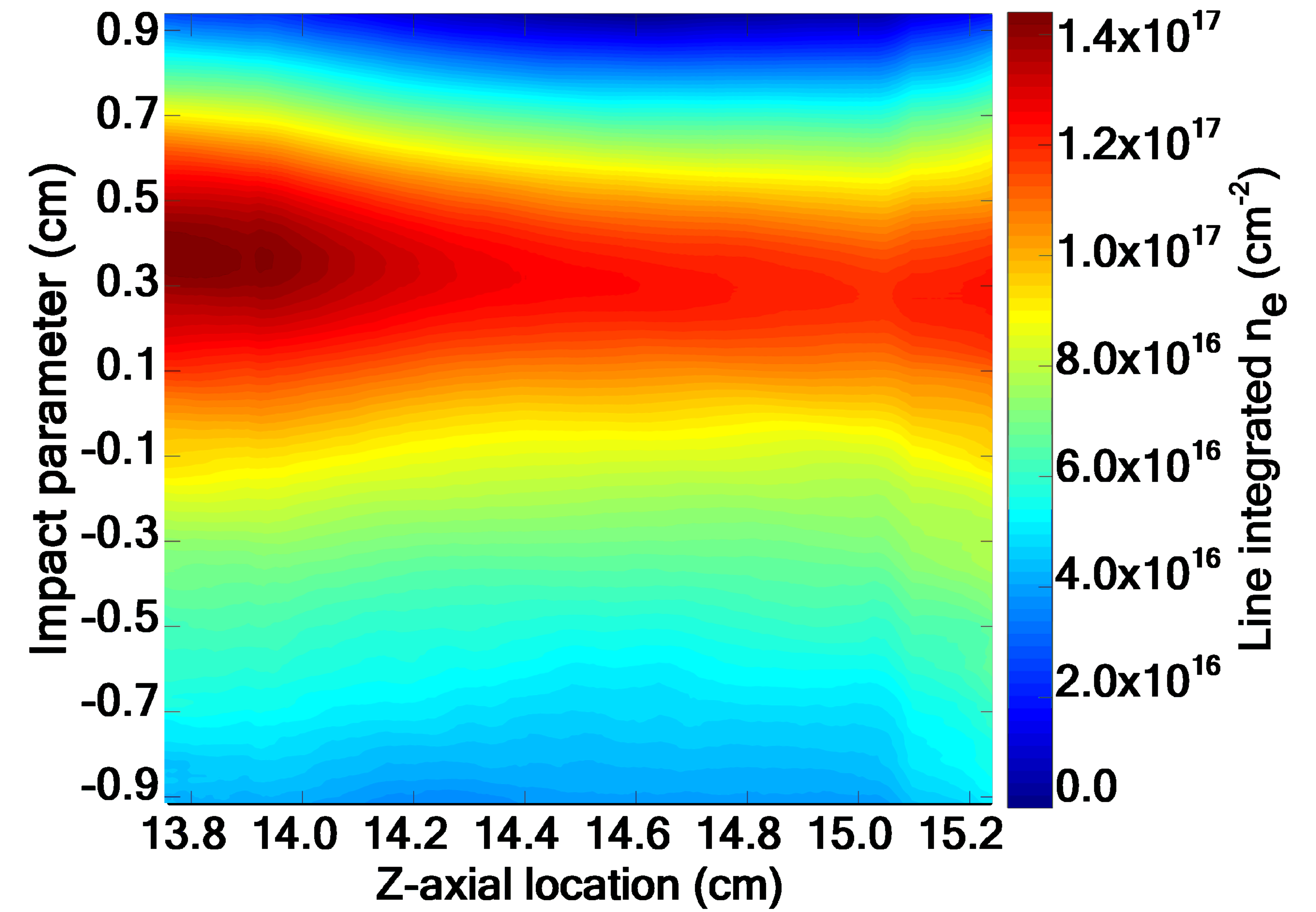} \label{fig:Line_DHI} }
\subfigure[] {\includegraphics[width=0.468\columnwidth{}]{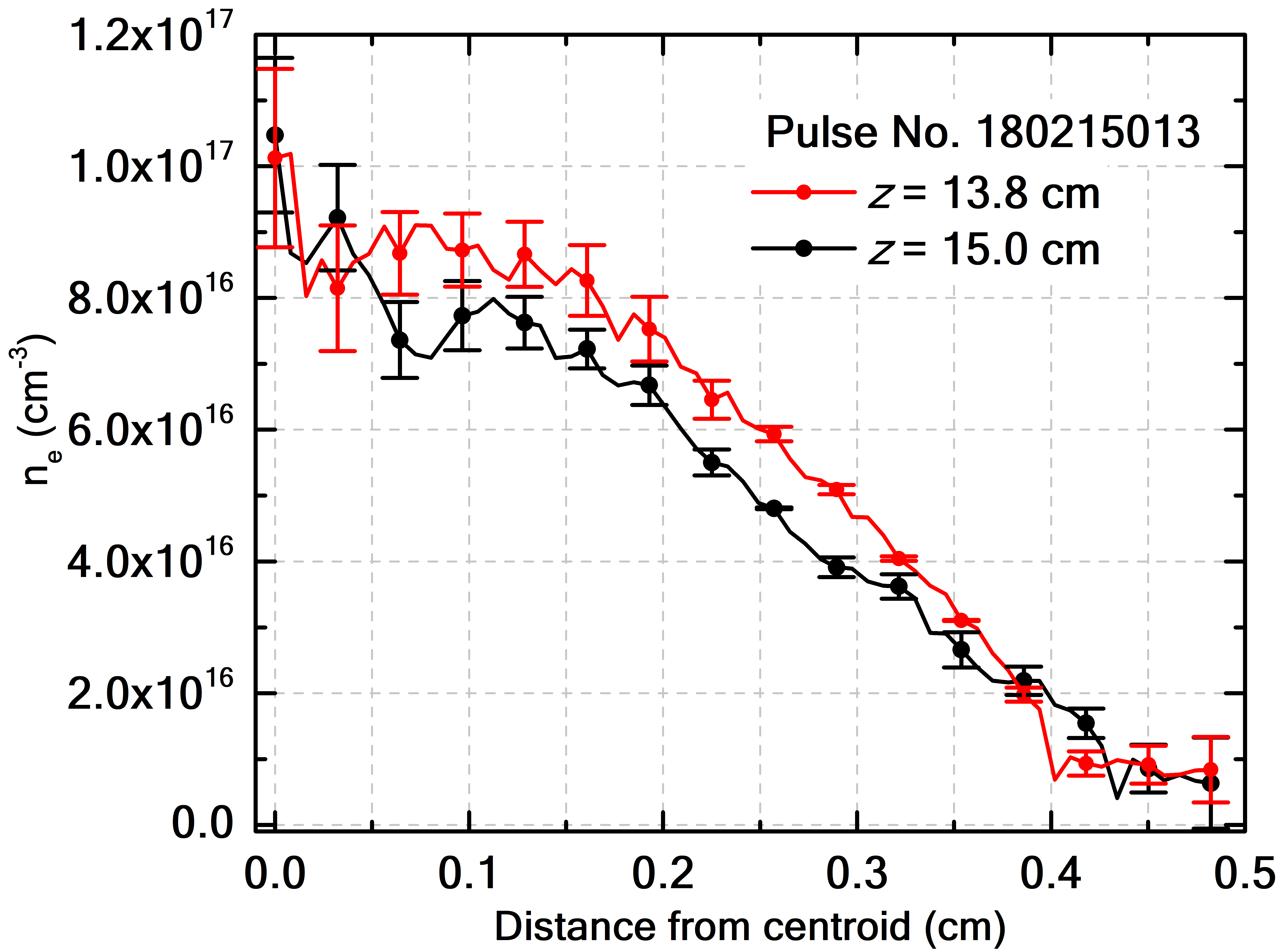}\label{fig:Radial_DHI} }
\caption{\protect\subref{fig:Line_DHI} Line-integrated plasma electron number density along $z$-axis and vertical impact parameter of a FuZE plasma pulse. A plasma pinch structure is observed.~\protect\subref{fig:Radial_DHI} Abel-inverted plasma electron particle density radial profiles at $z = 13.8$~cm and $z = 15.0$~cm with peak of approximately $1.1 \times 10^{17}$~cm$^{-3}$ and a pinch radius of approximately $0.3$~cm.}
\label{fig:DHI}
\end{figure}

The plasma density profile is measured using a digital holographic interferometer.\cite{jackson2006abel,ross2016digital} An image of the line-integrated plasma density profile is shown in Fig.~\ref{fig:Line_DHI}. The impact parameter indicates plasma vertical displacement from the $z$-axis. The Abel-inverted radial plasma density profiles at $z = 13.8$~cm and $z = 15.0$~cm are shown in Fig.~\ref{fig:Radial_DHI}. The pinch radius is approximately $0.3$~cm determined by HWHM (half width at half maximum) analysis. The plasma electron number density peaks at approximately $1.1 \times 10^{17}$~cm$^{-3}$.

\begin{figure}[htb]
\subfigure[] {\includegraphics[width=0.475\columnwidth{}]{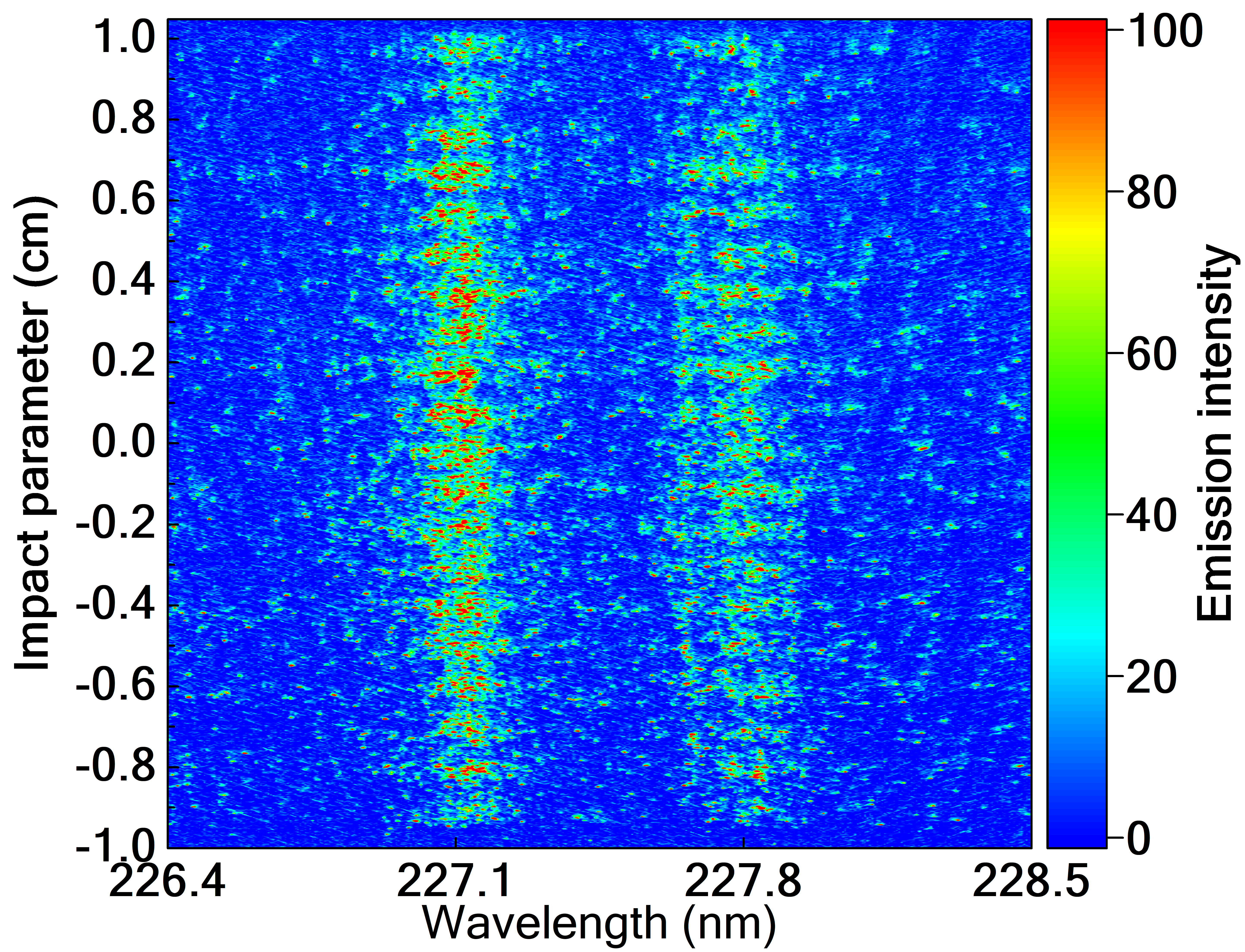}\label{fig:Img_ICCD} }
\subfigure[] {\includegraphics[width=0.495\columnwidth{}]{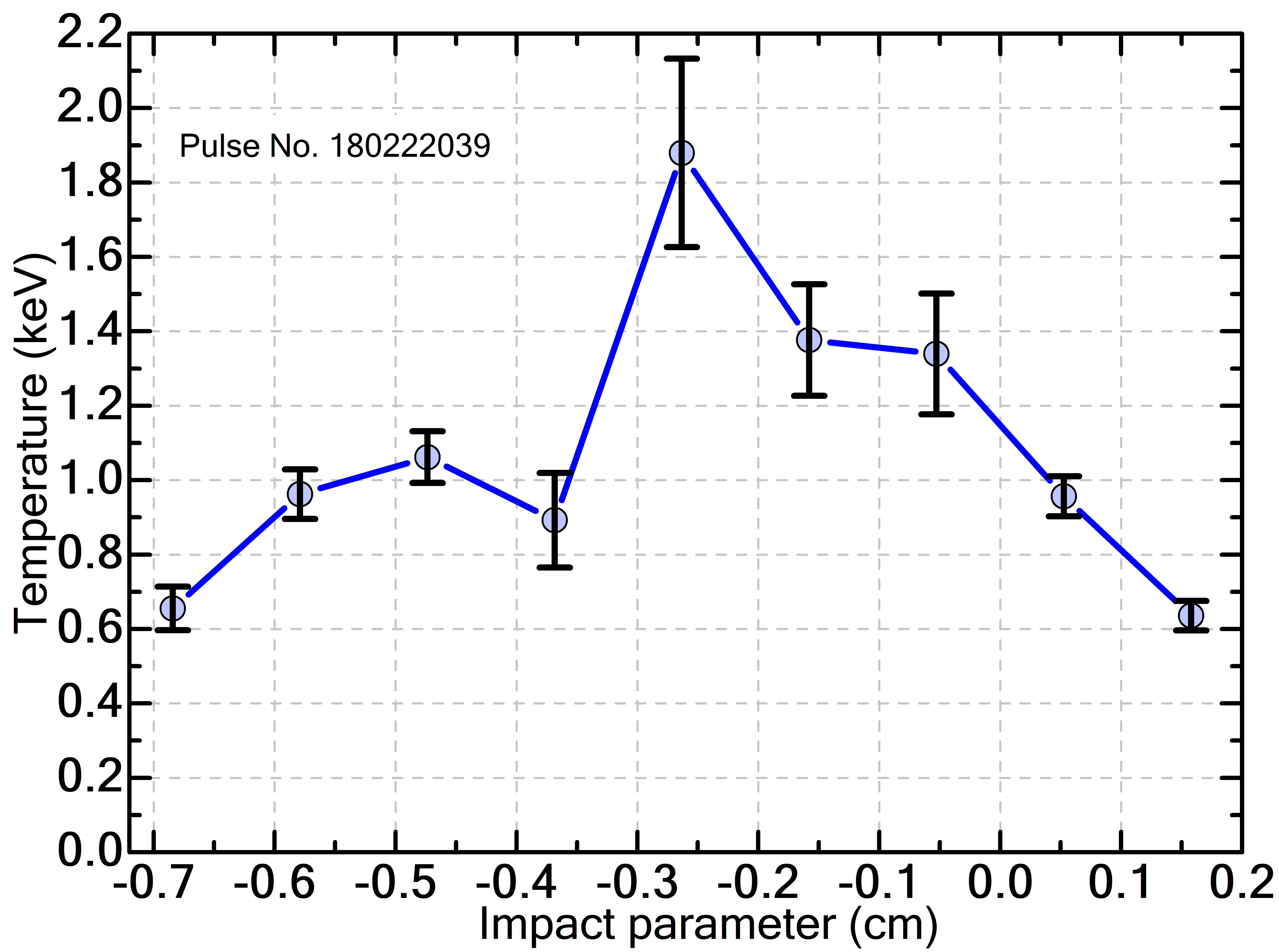}\label{fig:Radial_ICCD} }
\caption{\protect\subref{fig:Img_ICCD} Raw intensified charge-coupled device (ICCD) image shows the spatially resolved spectra of the carbon-V triplet lines ($227.1$ nm, $227.2$ nm and $227.8$ nm).~\protect\subref{fig:Radial_ICCD} Radial profile of the ion temperature indicates temperatures of $1-2$~keV. Included error bars are associated with the measurement accuracy and Gaussian-fit analysis. Temperatures from the plasma edge chords are not included since the emission intensity is too low for accurate fits.}
\label{fig:spt}
\end{figure}

Plasma temperature in the pinch is measured by ion Doppler spectroscopy based on spectra of impurity ions, such as carbon. Plasma densities are sufficiently high for collisions to thermalize the impurities and entrain their flow with the pinch plasma.\cite{shumlak2003sheared} The calculated ion-ion collisional equilibration time is approximately $100$~ns, which is a much smaller time scale than FuZE plasma quiescent time scale ($\approx 16~\mu$s). This indicates that the plasma reaches a local thermodynamic equilibrium, which justifies determining plasma ion temperature profiles indirectly by measuring the ion temperature of carbon. However, the equilibrium and the plasma parameters can evolve in time. During the quiescent period, carbon-V triplet lines are observed, as shown in Fig.~\ref{fig:Img_ICCD}. The radial profile of the ion temperature inferred from Doppler broadening is presented in Fig.~\ref{fig:Radial_ICCD}, indicating ion temperatures of $1-2$~keV. Note the impact parameter for this diagnostic indicates the plasma horizontal displacement from the $z$-axis.

Amplitudes of $m = 1$ fluctuations of magnetic field are obtained from 40 magnetic probes, distributed axially and azimuthally along the surface of the outer electrode in the assembly region. Data from these probes are Fourier analyzed to determine the time-dependent fluctuation levels of the $m = 1$ azimuthal mode $B_{1}\left(t,z\right)$ at each axial location, which are proportional to the radial displacement of the $Z$-pinch plasma current and indicate the centroid of the $Z$-pinch column. The average magnetic field $B_{0}\left(t,z\right)$ of all probes at a given axial location is used to normalize the Fourier mode data.\cite{golingo2007modeling} Measurement from the magnetic probes also provides evidence of axial plasma uniformity which has been presented elsewhere.\cite{shumlak2009equilibrium,shumlak2017increasing}

Neutron yields are measured using a cylindrical plastic scintillator, directly coupled to a fast PMT, that was calibrated at the High Flux Neutron Generator.\cite{ayllon2018design} Monte Carlo N-Particle (MCNP) calculations are used to determine the effective solid angle for the detectors and to correct for room-dependent neutron attenuation and scattering effects.\cite{mcnp6} The detector is located at $z = 15$~cm, and at a radial distance of $33$~cm from the $z$-axis.

\begin{figure}[htb]
\includegraphics[width=0.99\columnwidth{}]{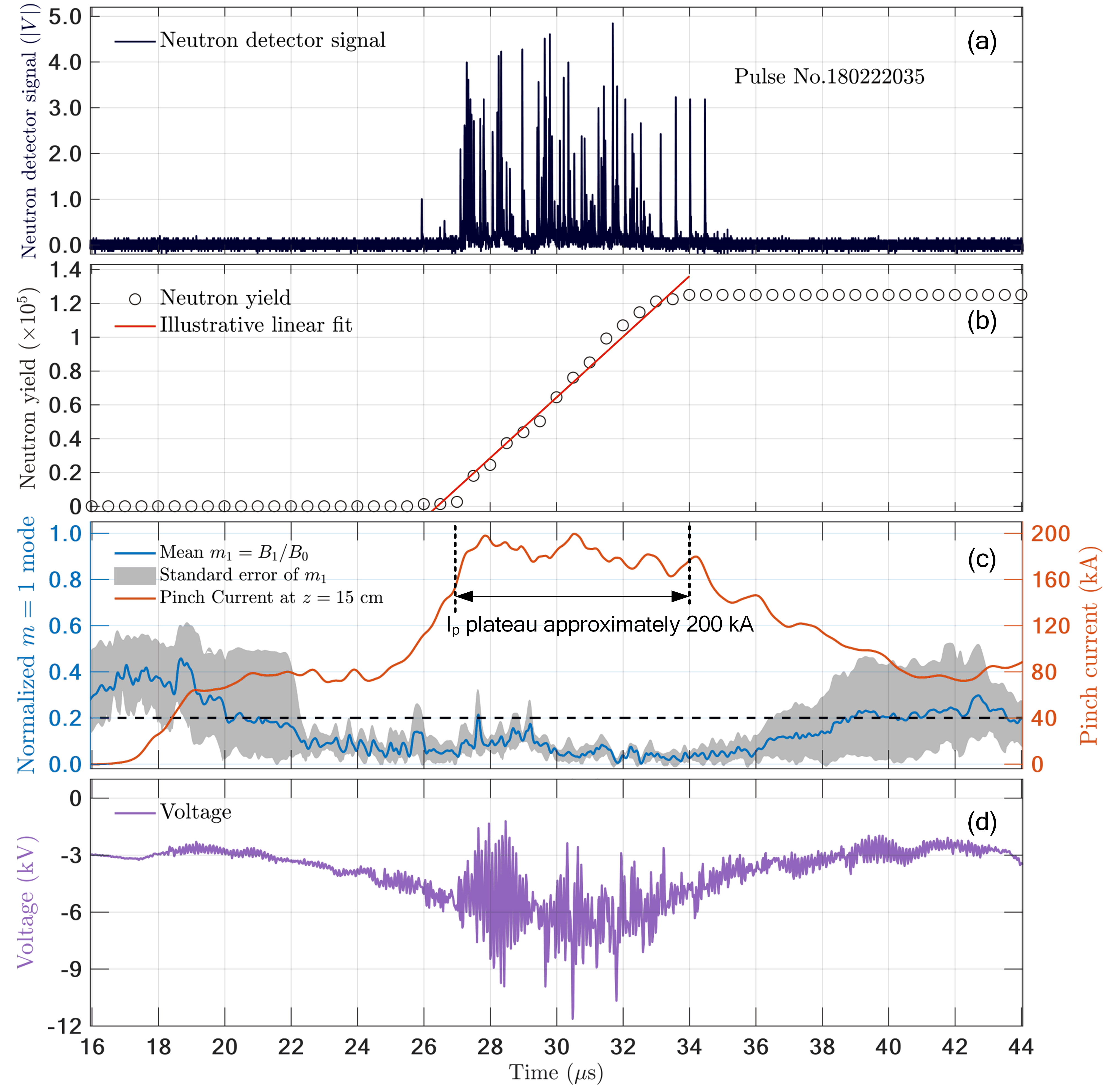}
\caption{(a): Characteristic signal is observed on the scintillator detector, showing neutron production during the quiescent period. For a $20\%$ deuterium$/80\%$ hydrogen gas mixture, the signal is sustained for approximately $5~\mu$s, which is 5000 times the $m = 1$ mode growth time. (b): Time-integrated neutron yield is calculated from measured scintillator detector signal. An illustrative linear fit plot is added to the figure, which indicates the neutron production rate is constant. (c): The normalized magnetic field fluctuation amplitude for the $m = 1$ mode, at locations $z$ = $5$~cm, $15$~cm, $25$~cm, $35$~cm, and $45$~cm, decreases below an empirical threshold of $0.2$ which indicates a quiescent plasma from $22~\mu$s to $38~\mu$s along the assembly region. The evolution of the plasma pinch current at $z = 15$~cm shows a high pinch current plateau of approximate $200$~kA during the sustained neutron production. (d): The voltage evolution shows no evidence of large voltage spikes during the quiescent period, indicating the absence of $m = 0$ instability.}
\label{fig:Bfluctuation}
\end{figure}

The neutron signal measured with a scintillator detector, the time-integrated neutron yield, analyzed data of magnetic fluctuation levels, plasma pinch current and pinch voltage signals are shown in Fig.~\ref{fig:Bfluctuation} for the case when injecting a $20\%$ deuterium$/80\%$ hydrogen partial pressure gas mixture. The plasma arrives at $z = 15$~cm at $17~\mu$s as indicated by plasma pinch current. Large magnetic fluctuations exist at the beginning of the $Z$-pinch formation. After the pinch forms at $22~\mu$s, the $m = 1$ fluctuations decrease to a low level, defining the beginning of an extended quiescent period which lasts for approximately $16~\mu$s for the data shown in Fig.~\ref{fig:Bfluctuation}. At $38~\mu$s, the fluctuation levels increase in amplitude, corresponding to the end of the quiescent period. During the quiescent period, a sustained scintillator signal is observed. The duration of the scintillator signal is approximately $5~\mu$s, coincident with the observed plateau of high pinch current ($\approx 200$~kA) and the absence of large voltage spikes.

\begin{figure}[htb]
\subfigure[] {\includegraphics[width=0.515\columnwidth{}]{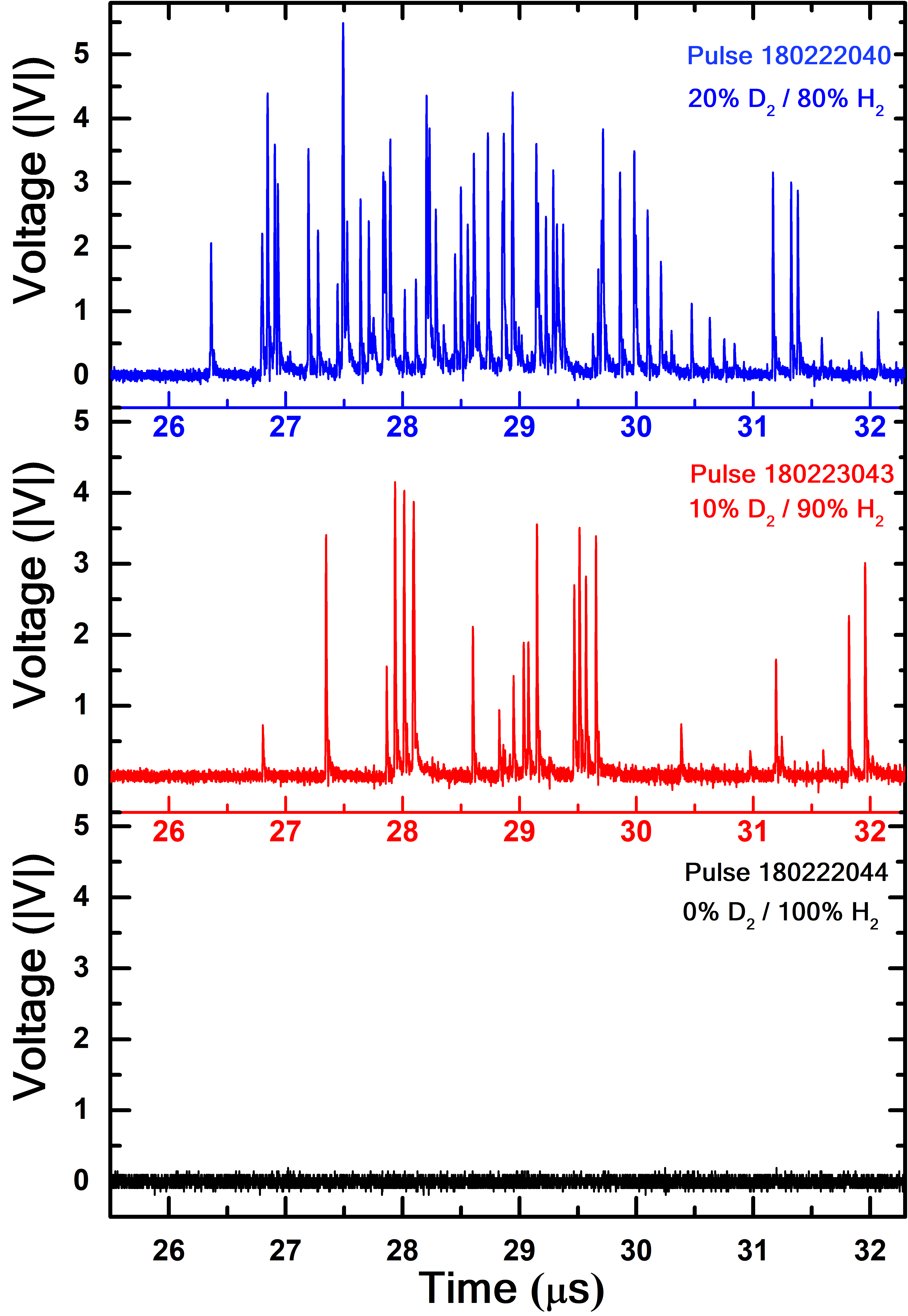}\label{fig:COM} }
\subfigure[] {\includegraphics[width=0.450\columnwidth{}]{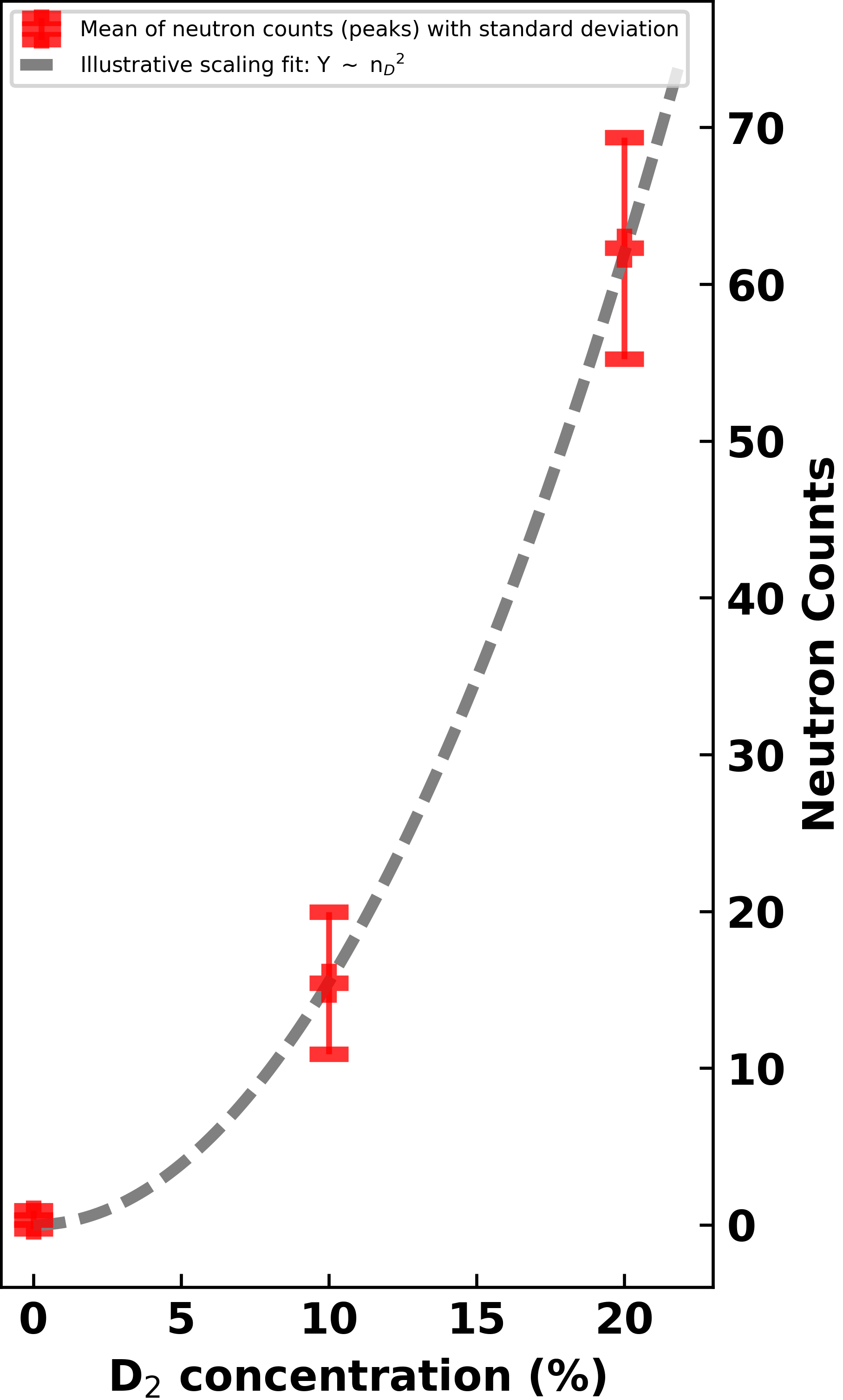}\label{fig:Pre} }
\caption{\protect\subref{fig:COM} Comparison of plastic scintillator detector signals during the quiescent period of a $Z$-pinch plasma for three different deuterium concentrations: 20\%, 10\%, and 0\%. For non-zero deuterium concentrations, clearly detectable neutrons signals are observed. No neutron signal is detected in the absence of deuterium fill gas. \protect\subref{fig:Pre} The average number of neutron counts observed at three different gas mixtures agree with an expected ${n_{D}}^2$ scaling relation. Error bars indicate the uncertainty in the neutron counts and the shot-to-shot variation.}
\label{fig:NDcom}
\end{figure}

To further investigate the origin of scintillator detector signals, experiments were conducted using three different deuterium partial pressure concentrations: 20\%, 10\%, and 0\%. As shown in Fig.~\ref{fig:COM}, signals are observed on the plastic scintillator detector with non-zero deuterium mixtures. However, for the 0\% deuterium case, no signal is observed. This result strongly indicates that the measured scintillator detector signals are from neutron emissions, and are not due to X-rays. Tens of pulses for each deuterium concentration setting are recorded, and all produced comparable levels of neutron yields as denoted by the error bars in Fig.~\ref{fig:Pre}. Statistical analysis of the data in Fig.~\ref{fig:Pre} shows an average yield of $\left ( 1.25\pm 0.45 \right )\times 10^{5}$ neutrons per pulse for $20\%$ deuterium$/80\%$ hydrogen gas partial pressure mixture. 

Mitigating the growth and development of the $m = 0$ sausage mode and $m = 1$ kink mode is critical to $Z$-pinch performance. As shown in Fig.~\ref{fig:Bfluctuation}, during the neutron emission, the voltage signal shows no evidence of large voltage spikes, indicating the absence of $m = 0$ instabilities.\cite{sethian1987enhanced,lebedev1998coronal} In practical, the occurrence of an $m = 0$ instability is indicated by a single voltage spike above $20$~kV in FuZE. Meanwhile the decrease of $m = 1$ fluctuations to a low level demonstrates the suppression of $m = 1$ instabilities as well. The theoretical $m = 1$ mode growth time for a static $Z$-pinch plasma is approximately $\left(kV_{A}\right)^{-1}$.\cite{shu1995sheared} For the presented FuZE plasma parameters, the calculated Alfv$\acute{\mathrm{e}}$n velocity is $\approx 8.77 \times 10^5$ m/s. Assuming an axial wavelength equal to the plasma diameter, the theoretical growth time is approximately 1~ns. The $5~\mu$s time span of the observed neutron emission signal is 5000 instability growth times. This sustained neutron emission, coincident with the lack of $m = 0$ and $m = 1$ instabilities, suggests a thermonuclear fusion process may be responsible for the neutron observation, not a beam-target fusion process, where neutrons are generated from instabilities and the neutron duration lasts for a few tens or hundreds of nanoseconds.\cite{haines2011review,klir2015efficient}

For the three different deuterium concentrations shown in Fig.~\ref{fig:Pre}, neutron production results follow the expected ${n_{D}}^2$ dependence,\cite{velikovich2007z} which provides additional evidence of possible thermonuclear fusion with the deuterium mixture plasmas. However, further investigation is needed to better characterize the energy spectrum of the observed neutrons.

The thermonuclear neutron yield from a $Z$-pinch plasma column is\cite{velikovich2007z}
\begin{equation} \label{eq:1}
Y=\frac{1}{2}\int n_{D}(r)^{2} \left \langle \sigma \nu\right \rangle_{T} \cdot 2\pi rdr \cdot l\cdot\tau,
\end{equation}
where $n_{D}(r)$ is the radially-dependent deuterium ion number density, $\left \langle \sigma \nu\right \rangle_{T}$ is the ion-temperature-dependent D(d,n)$^3$He fusion reaction
rate parameter,\cite{bosch1992improved}, $r$ and $l$ are the $Z$-pinch radius and length, and $\tau$ is the neutron-emission pulse length.

Assuming a Bennett-type equilibrium density profile $n_i(r) = n_{i0}/[1+(r/r_0)^2]$ \cite{bennett1934magnetically}, $n_{D} = 0.2n_{i}$ for a 20\% deuterium concentration, peak ion number density $n_{i0}$ = $10^{17}$ cm$^{-3}$, pinch radius $r_0$ of $0.3$~cm, pinch length of $50$~cm, a neutron-emission pulse length of $5~\mu$s and the D(d,n)$^3$He fusion reaction rate parameter, the experimentally-measured neutron yield $1.25\times 10^5$ neutrons/pulse, with 36\% uncertainty, gives a calculated plasma ion temperature of $1.1-1.3$~keV based on Eq.~(\ref{eq:1}). This calculated result agrees well with the measured ion temperature shown in Fig.~\ref{fig:Radial_ICCD}. This indicates that the measured neutron yield $1.25\times 10^5$ neutrons/pulse is bracketed by the independent experimental plasma parameter measurements including measurement uncertainties.

Considering a steady-state isotropic $50$~cm line neutron source emitting at rate of $10^5$ neutrons over $5~\mu$s, Poisson statistics analysis shows a theoretical waiting time between two neutron counts is $50.0\pm7.1$~ns (with $95\%$ confidence) at the detector's experimental location. The measured neutron-emission pulse gives a waiting time of $62.7\pm7.9$~ns. The observed time-variations of neutron signals is within statistical expectations of a steady-state isotropic neutron source. Fig.~\ref{fig:Bfluctuation} plots the time evolution of the total neutron yield. As expected for a steady-state source, the data are well-described by a linear fit.

Neutron emission is observed during the plasma quiescent period but only during the $5~\mu$s high-current plateau since the fusion reaction rate strongly depends on plasma density and temperature, as indicated in Eq.~(\ref{eq:1}). The plasma density increases with pinch current as a result of compression.\cite{shumlak2017increasing,shumlak2012sheared} The fusion reaction rate parameter strongly depends on plasma temperature, $\left \langle \sigma \nu\right \rangle_{T}\propto T^4$ in the observed temperature range.\cite{bosch1992improved} According to the Bennett relation for Z-pinch equilibrium, plasma temperature $T$ is proportional to the square of the pinch current.\cite{bennett1934magnetically} The neutron production rate is then expected to scale as the pinch current to a power greater than eight.

Between the start of the quiescent period and the plateau of high pinch current, the plasma current increases by a factor of $2.5$ from $80$~kA to $200$~kA. According to the scaling, the rate of neutron production is expected to increase by over three orders of magnitude. This scaling offers a possible explanation why the observed neutron emission signals are coincident with elevated pinch currents and plasma stability.

In summary, employing the SFS $Z$-pinch concept, FuZE has achieved equilibrium-stabilized plasma with fusion-relevant parameters of $10^{17}$~cm$^{-3}$ number density, $1-2$~keV temperature, $0.3$~cm pinch radius, and long-lived quiescent periods of approximately $16$~$\mu$s on a scale that facilitates diagnostic measurements. The demonstration of sustained neutron production lasting approximately $5$~$\mu$s, thousands of the theoretical $m = 1$ mode growth time, the absence of $m = 0$ and $m = 1$ instabilities during neutron production, and the observation of neutron yield scaling with ${n_{D}}^2$. indicate consistency with a thermonuclear fusion process. The measured neutron yields are approximately $10^5$ neutrons/pulse, consistent with theoretical expectations for the measured plasma parameters and within the statistical expectations of a steady-state line neutron source. Although further investigation is needed to better characterize the energy spectrum of the observed neutrons, the results presented in this Letter provide a compelling argument for continued pursuit of SFS $Z$-pinch concept towards high-energy-density physics (HEDP) and fusion physics.

This work is funded in part by the Advanced Research Projects Agency - Energy (ARPA-E), U.S. Department of Energy, under Award Number DE-AR-0000571 and LLNL Contract DE-AC52-07NA27344. U. Shumlak gratefully acknowledges support of the Erna and Jakob Michael Visiting Professorship at the Weizmann Institute of Science and as a Faculty Scholar at the Lawrence Livermore National Laboratory.

\end{document}